\documentclass[%
 reprint,
 amsmath,amssymb,
 aps,
]{revtex4-2}

\usepackage[T1]{fontenc} \usepackage[latin1]{inputenc}
\usepackage{amsmath,color,ulem} \usepackage{graphicx}
\usepackage{epsf,bm} \usepackage{amssymb}

\usepackage{tabulary}
\newcolumntype{K}[1]{>{\centering\arraybackslash}p{#1}}

\usepackage{multirow}

\begin{document}

\title{Crossover Scaling of Binder Cumulant and its application in Non-reciprocal Sandpiles}
\author{Wei Zhong$^1$} 
 
\author{Youjin Deng$^{2,3}$} 
\email{yjdeng@ustc.edu.cn} 
 
\affiliation{
 $^1$ Minjiang Collaborative Center for Theoretical Physics, College of Physics and Electronic Information Engineering,
Minjiang University, Fuzhou 350108, P. R. China.\\
$^2$ Hefei National Laboratory for Physical Sciences at the Microscale and Department of Modern Physics, University of Science and Technology of China, Hefei 230026, China\\
$^3$ Hefei National Laboratory, University of Science and Technology of China, Hefei 230088, China
}

\date{\today} 

\begin{abstract} 

In this letter, we unveil a robust, pre-asymptotic scaling regime for the Binder cumulant $U_L$, a central finite-size scaling tool, demonstrating $U_L\sim N^{-1} |t|^{-d\nu}$ (disordered phase) and $\frac{2}{3}-U_L\sim N^{-1} |t|^{-d\nu}$ (ordered phase), with $t$ being the reduced control parameter, and $N$, $d$, $\nu$ represent the total number of sites, the dimensionality, and correlation length exponent, respectively. Leveraging this result, we resolve a fundamental question on the stability of universality classes under the breaking of microscopic reciprocity. For the conserved Manna sandpile, we show that reciprocal biases preserve its universality class, merely shifting the critical point. In striking contrast, any non-reciprocal interaction acts as a relevant perturbation, decisively driving the system's critical exponents to flow from their non-mean-field values towards the mean-field related ones. This flow
 establishes non-reciprocity as a generic mechanism inducing mean-field criticality in conserved, non-equilibrium systems.
\end{abstract}

\maketitle

{\it Introduction--}
The classification of continuous phase transitions through universal critical exponents represents a cornerstone of statistical physics. In computational studies, finite-size scaling (FSS) serves as the indispensable methodology for extracting these exponents from systems of finite extent \cite{newman1999monte,ferrenberg1991critical}. A central tool within FSS is the Binder cumulant $U_L$, a dimensionless ratio constructed from the moments of the order parameter distribution \cite{binder1981finite,binder1992monte}. Its principal utility arises from the intersection method, where $U_L(T)$ curves for different linear sizes $L$ cross at a common point, providing a numerically precise estimate of the critical point\cite{binder1981critical,fisher1982scaling}. This technique has found ubiquitous application. However, a comprehensive theoretical framework for the {\it complete scaling function} of $U_L$, particularly in low-dimensional systems and in the region surrounding\verb|--|rather than\verb|--|the critical point, remains unclear. 

Conventional analyses predominantly rely on its asymptotic value at critical point or on simple FSS ansatzes valid only in the immediate vicinity of criticality \cite{blote1995ising}, {\it e.g.},
\begin{equation}
U_L(x)\approx U_c+a_1 x+a_2 x^2 + ... \quad \text{with} \quad x=t L^{1/\nu},
\end{equation}
where $a_n$ are constants, $U_c$ represents the value of the Binder cumulant at the critical point, $x$ is on order of $\mathcal{O}(1)$, and $t\sim L^{-1/\nu}$ vanished algebraically as $L$ increases. The behavior in the pre-asymptotic finite-size window, where the correlation length $\xi \sim |t|^{-\nu}$
is larger than $1$ but significantly smaller than the side length $L$, is less understood. Elucidating a robust scaling law for $U_L$ in this regime would furnish a powerful, complementary diagnostic for critical phenomena, especially where strong corrections complicate traditional analysis.

This methodological advancement is acutely relevant for unraveling the critical properties of complex non-equilibrium systems. A prototypical case is the conserved Manna sandpile \cite{bak1987self,dhar1999abelian,manna1991two,vespignani1998driving,
basu2012fixed,wiese2024hyperuniformity,manna2025describing}. Despite its status as a distinct universality class, numerical determinations of its critical exponents have exhibited notable variation, a discrepancy widely attributed to substantial non-analytic finite-size corrections and slow crossover effects that mask the true asymptotic behavior \cite{lubeck2004universal,hinrichsen2000non}. This inherent difficulty in achieving precise convergence underscores the need for refined scaling approaches. 

The challenge extends further when considering perturbations that break fundamental microscopic symmetries. Of particular contemporary interest are non-reciprocal interactions, where the coupling from site $i$ to $j$ differs from that of $j$ to $i$. Such directed couplings are intrinsic to a broad spectrum of active, driven, and non-Hermitian systems \cite{weber2014defect,fily2012athermal,bergholtz2021exceptional,
fruchart2021non,duan2025phase,martin2025transition,belyansky2025phase,
avni2025nonreciprocal}. Introducing non-reciprocity into a well-established critical system like the conserved Manna sandpile poses a profound question: does it keep the universality class, or does it fundamentally destabilize the original fixed point?

In this Letter, we bridge this methodological gap and answer the posed physical question. First, we report a novel, two-sided scaling law for the Binder cumulant in the pre-asymptotic regime, validated across models and at the mean-field limit. This provides a direct spectroscopic tool for measuring the correlation length exponent $\nu$. Second, we apply this tool to the {\it non-reciprocal conserved Manna sandpile}. We demonstrate that while reciprocal hopping probability preserves the Manna universality class, any finite non-reciprocal interaction drives a rapid renormalization group flow to mean-field criticality. This finding is robust across different bias patterns 
establishing non-reciprocity as a generic mechanism for mean-field behavior in conserved non-equilibrium systems.

{\it Pre-asymptotic Scaling behavior for the Binder Cumulant--} The precise extraction of critical exponents from finite systems hinges on a deep understanding of finite-size scaling (FSS) functions. While the Binder cumulant, $U_L$, is a standard tool for locating critical points\cite{binder1981finite,binder1981critical}, its full scaling function-particularly its behavior in the pre-asymptotic regime where the correlation length $\xi$ is significant yet $\xi\ll L$-remains less charted. This regime is critical for analyzing systems with strong scaling corrections. Here, we establish a universal, two-sided scaling framework for $U_L$ in the pre-asymptotic regime, directly linking its behavior to the exponent $\nu$. We illustrate this framework using the paradigmatic two-dimensional (2D) Ising model.

We perform high-precision simulations of the 2D ferromagnetic Ising model on an $L\times L$ square lattice, using the Wolff cluster algorithm \cite{wolff1989collective} to access system sizes $L$ up to $256$. The Binder cumulant is computed from the magnetization moments\cite{binder1981finite}:
\begin{equation}
U_L=1-\frac{\langle m^4\rangle}{3 \langle m^2\rangle^2},
\end{equation}
where the order parameter $m=\sum_{i=1}^{N} s_i/N$, with $s_i$ be the spin value at size $i$ and $N=L\times L$ is the total sites.

\begin{figure}[htp]
	\centering
	\includegraphics[width=0.95\linewidth]{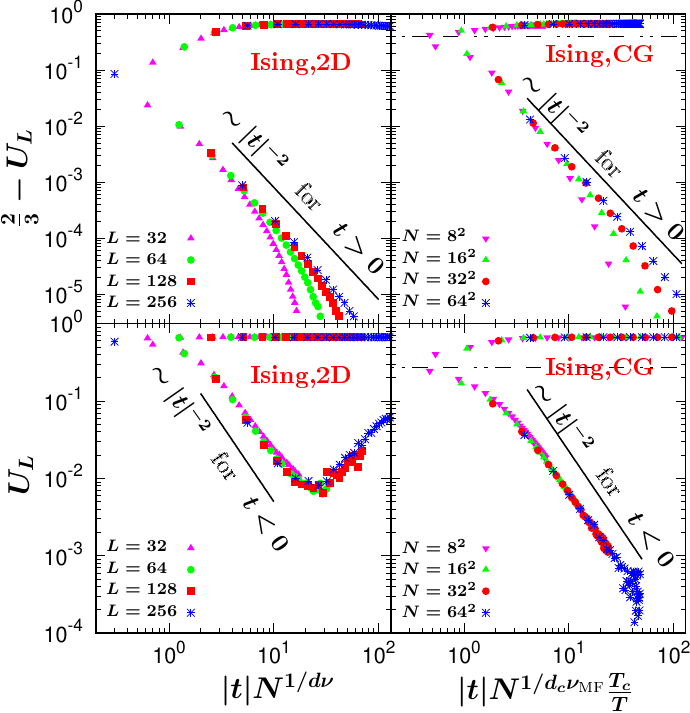} 
	\caption{Discovery and validation of the pre-asymptotic Binder cumulant scaling in the 2D Ising model and Ising model on complete graph (CG). The results indicate that  $U_L\sim |t|^{- d \nu}$ for $T>T_c$ and $\frac{2}{3}-U_L \sim |t|^{-d \nu}$ for $T<T_c$, where $d\nu=2$ for the 2D Ising model and $d_c \nu_{\text{MF}}=2$ for Ising model on complete graph (mean-field results). The dash lines represent the mean-field values of $\frac{2}{3}-U_{L\rightarrow \infty}$ and $U_{L\rightarrow \infty}$ at the critical point (see Eq. (8)).
	}
	\label{fig1}
\end{figure}

Conventional FSS analysis focuses on the size-dependent value $U_L(T_c)$ or the intersection of $U_L(T)$ curves to locate $T_c$. Our analysis targets the pre-asymptotic region where $\xi\sim |t|^{-\nu}$ (with $t=1-\frac{T}{T_c}$) is large but distinctly smaller than $L$. We discover that $U_L$ obeys 
scaling laws on either side of the critical point (FIG. \ref{fig1}, left). In the high-temperature (disordered) phase, $U_L$ itself approaches zero from above, following a power law:
\begin{equation}
U_L\sim  N^{-1}|t|^{-d \nu}, \quad{\text for} \quad t<0, 1\ll \xi\ll L.
\end{equation}
In the low-temperature (ordered) phase, $U_L$ approaches its saturation limit of $\frac{2}{3}$ from below, with the deviation exhibiting a complementary power law: 
\begin{equation}
\frac{2}{3}-U_L\sim  N^{-1}|t|^{-d \nu}, \quad {\text for} \quad t>0, 1\ll \xi\ll L.
\end{equation}
For the 2D Ising model, we have $d \nu = 2$.

 Our results indicate that the scaling function $U_L(t)$ has distinct asymptotic limits: $U_L \sim N^{-1} |t|^{-d\nu}$ for $t<0$ (disordered side), and $\frac{2}{3}-U_L \sim N^{-1} |t|^{-d\nu}$ for $t>0$ (ordered side).  
Which means that we have $U_L$ or $\frac{2}{3}-U_L \propto V_{\text{eff}}^{-1}$, where $V_{\text{eff}}=\left[\frac{L}{|T-T_c|^{-\nu}}\right]^d=\left[\frac{L}{\xi}\right]^d$ is the effective volume. 
It reveals that the measured power laws are not artifacts but fundamental features of the universal crossover from critical to non-critical fluctuations. 
 
To confirm the universality of this pre-asymptotic scaling beyond the 2D Ising universality class, we have performed supplementary analyses on other models. Notably, measurements on the 3D Ising model and the 2D site percolation model yield results fully consistent with its known exponent product $d\nu$ ({\it Supplement Materials, S1}). This confirms that the scaling form is not model-specific but reflects a general property of the Binder cumulant in the pre-asymptotic regime.
 
The generality of this scaling is further confirmed at the upper critical dimension. For the mean-field Ising model, supposing the free energy is given by
\begin{equation}
f(m)=a/2 m^2+b/4 m^4
\end{equation}
where $a=a_0(T-T_c)$ and $b$ are two parameters. Then the moments of the magnetization expresses as 
\begin{equation}
\langle m^k\rangle = \frac{1}{Z} \int_{-\infty}^{\infty} dm m^k e^{-N \beta f(m)},
\end{equation}
where $Z= \int_{-\infty}^{\infty} dm e^{-N \beta f(m)}$ is the partition function, $\beta=1/(k_B T)$ is the inverse temperature, and $k_B$ is the Boltzmann constant. 

By using the saddle-point expansion method ({\it Supplement Materials, S2}), we have 
\begin{align}
    \frac{2}{3}-U_{L\rightarrow \infty} &\approx  \frac{2 k_B T b}{3 N a_0^2 T_c^2 |t|^2} \quad for \quad t>0,\\
    U_{L\rightarrow \infty} &\approx 0.27044 \quad for \quad t=0,\\
         U_{L\rightarrow \infty} & \approx  \frac{2 k_B T b}{N a_0^2  T_c^2 |t|^2} \quad for \quad t<0.
\end{align}

Note that within the Bragg-Williams approximation that $b\propto T$ for the Ising model \cite{wipf2021statistical}. When $N\rightarrow \infty$ and $T$ near $T_c$, we can treat the temperature as a constant $T\approx T_c$, then the mean-field analytical calculation predicts both $U_{L\rightarrow \infty} \sim  N^{-1}|t|^{-d_{c} \nu_{\text{MF}}}$ for $t>0$ and $\frac{2}{3}-U_{L\rightarrow \infty} \sim  N^{-1}|t|^{-d_{c} \nu_{\text{MF}}}$ for $t<0$ ($d_{c} \nu_{\text{MF}}=2$ \cite{binder1997applications}). Our Monte Carlo simulations on complete graph the spin number up to $N = 64^2$ spins confirm these predictions (FIG. \ref{fig1}, right).



This pre-asymptotic scaling framework for $U_L$ offers significant advantages. As noted earlier, substantial non-analytic finite-size corrections and slow crossover effects \cite{lubeck2004universal,hinrichsen2000non} introduce considerable variations in numerical estimates of critical exponents for non-equilibrium systems. For instance, even for the one-dimensional conserved Manna sandpile\verb|--|one of the simplest models exhibiting a non-equilibrium phase transition\verb|--|was believed to belong to the universality class of $(1+1)$-dimensional directed percolation until 2010s \cite{basu2012fixed,kwon2015critical}. The pre-asymptotic scaling of the Binder cumulant provides a complementary and robust route for extracting $\nu$, mitigating the sensitivity to corrections to scaling and uncertainties in the critical point that affect methods precisely tuned to $T_c$ \cite{ferrenberg1991critical,blote1995ising}.

Most importantly for our purpose, this framework serves as a highly sensitive diagnostic for crossover phenomena. When perturbations, like the non-reciprocal interactions in the 2D Manna sandpiles, are introduced that break fundamental microscopic symmetries, it remains unclear whether the universality class of the system is preserved. A clean power-law behavior in a plot of $\frac{2}{3}-U_L$ versus $|t|$ confirms a stable effective exponent. In contrast, systematic curvature in such a plot provides unambiguous evidence of a flowing effective exponent $\nu_{\text{eff}}$, signaling that the system is undergoing a renormalization group flow between fixed points.

\begin{figure}[htp]
	\centering
	\includegraphics[width=0.48\linewidth]{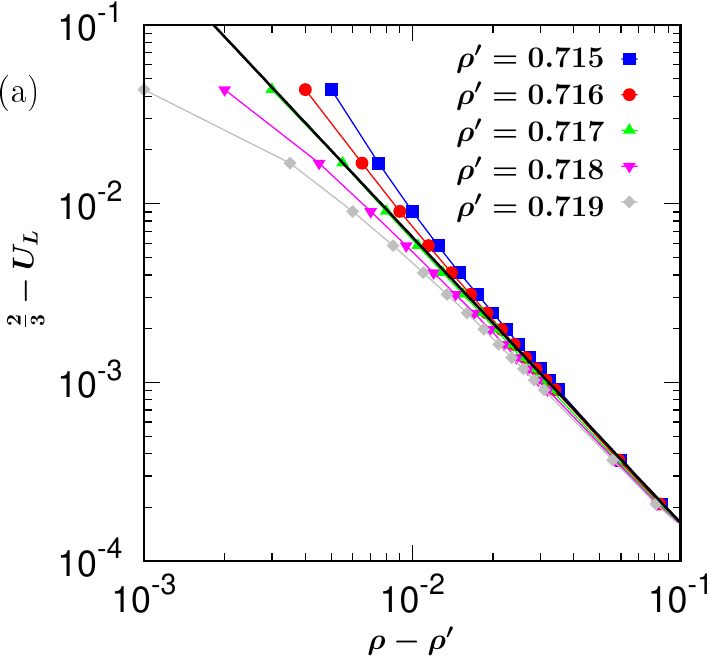} 
	\hspace{1mm}
	\includegraphics[width=0.48\linewidth]{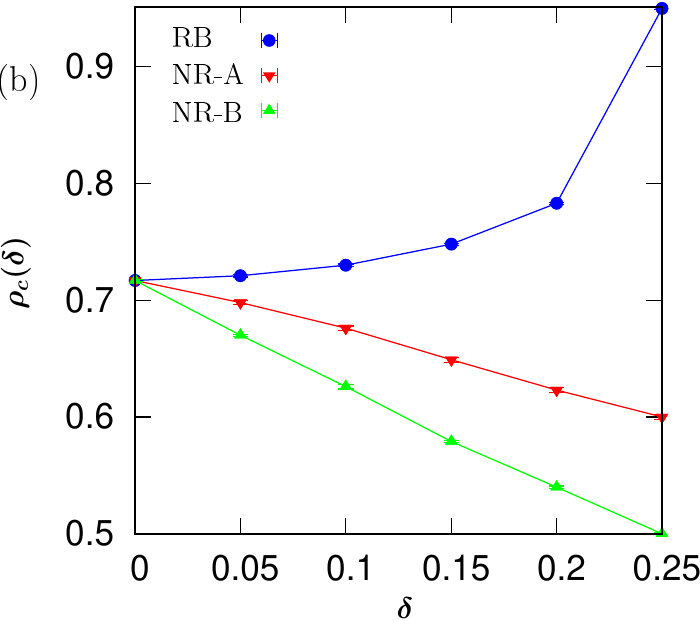}
	\caption{ (a) The log-log plot of $\frac{2}{3}-U_L$ {\it v.s.} $\rho-\rho'$ for $L=256$. By adjusting $\rho'$, we obtain the critical density $\rho_c$ when $\frac{2}{3}-U_L$ exhibits power-law behavior verse $\rho-\rho'$. (b) Critical density $\rho_c$ versus bias strength $\delta$ for Reciprocal Bias (RB, blue bullet), Non-reciprocal Bias A (NR-A, red down triangle), and Non-reciprocal Bias A (NR-B, green up triangle).}
	\label{manna_tc}
\end{figure}

As an important example, in Fig. \ref{manna_tc}(a), we demonstrate the determination of the critical point for the standard 2D Manna sandpile by plotting $\frac{2}{3}-U_L$ against $\rho-\rho'$, where $\rho$ is the density of grains. By adjusting $\rho'$, we obtain a clean power-law behavior of $\frac{2}{3}-U_L$ at an appropriate value $\rho' = \rho_c$. This yields the precise estimates $\rho_c \approx 0.717$ and the correlation length exponent $\nu \approx 0.80$, which are consistent with existing results \cite{karmakar2004directed,henkel2008}

In summary, we have established a unified, two-sided scaling framework for the Binder cumulant in the pre-asymptotic regime: $U_L \sim N^{-1} |t|^{-d \nu}$ ($t<0$) and $\frac{2}{3}-U_L \sim N^{-1} |t|^{-d \nu}$ ($t>0$). For absorbing phase transitions such as the Manna sandpile, the latter relation provides a powerful and direct spectroscopic tool for measuring $\nu$ and diagnosing crossover physics. We now apply this tool to unravel the critical behavior of the non-reciprocal conserved Manna sandpile.

{\it Criticality of the Non-reciprocal Conserved Manna Sandpile--} Armed with the new scaling tool established in last section, we now investigate the core physical question: how does microscopic non-reciprocity alter universality in a well-established non-equilibrium critical system? 

We focus on the 2D conserved Manna sandpile \cite{dhar1999abelian,manna1991two,vespignani1998driving}, which is a variant of the original Manna model in which the total number of grains (energy) is kept constant, rather than being injected at a slow rate. The system is defined on an $L\times L$ square lattice, and each site holds a non-negative integer number of grains. Initially, the lattice is populated with a fixed total mass distributed randomly. At each time step, an active site $i$, which contains at least two grains, {i.e.}, $z_i\geq 2$, is selected uniformly at random. That site topples by removing two grains from it and transferring one grain to each of two distinct neighbors chosen uniformly at random.


The innovation is the introduction of biased toppling probabilities. Let $q_{\text{right}}$, $q_{\text{left}}$, $q_{\text{up}}$, $q_{\text{down}}$ denote the probabilities for a transferred particle to move right, left, up, or down, respectively ($q_{\text{right}}+q_{\text{left}}+q_{\text{up}}+q_{\text{down}}=1$). We study three structured bias protocols controlled by $\delta \in [0,0.25]$:

{\bf 1 Reciprocal Bias (RB)}: $q_{\text{right}}=q_{\text{left}}=0.25+\delta$, $q_{\text{up}}=q_{\text{down}}=0.25-\delta$. This preserves spatial inversion symmetry. 

{\bf 2 Non-reciprocal Bias A (NR-A)}: $q_{\text{right}}=q_{\text{left}}=q_{\text{up}}=0.25+\delta/3$, $q_{\text{down}}=0.25-\delta$.

{\bf 3 Non-reciprocal Bias B (NR-B)}: $q_{\text{right}}=q_{\text{up}}=0.25+\delta$, $q_{\text{left}}=q_{\text{down}}=0.25-\delta$. 

Protocols NR-A and NR-B break inversion symmetry in distinct ways, establishing a global non-reciprocal current. The standard isotropic model is $\delta=0$.

The order parameter is the density of active sites, $m=\langle \rho_a\rangle$. Following Sec. II, we perform finite-size scaling ($L$ up to $256$). Note that the absorbing phase ($t>0$, here $t=1-\frac{\rho}{\rho_c}$) has a vanishing order parameter, making the standard definition of $U_L$ trivial (effectively zero) and its high-temperature scaling law inapplicable for exponent extraction. Therefore, our analysis of the non-reciprocal sandpile will exclusively leverage the robust scaling of $\frac{2}{3}-U_L$ in the active phase ($t<0$).

\begin{figure}[htp]
	\centering
	\includegraphics[width=0.48\linewidth]{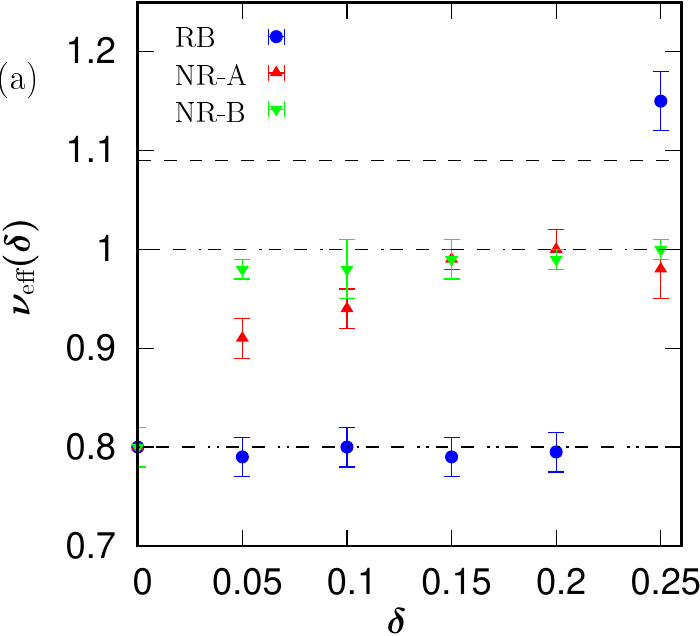} \hspace{1mm}
	\includegraphics[width=0.48\linewidth]{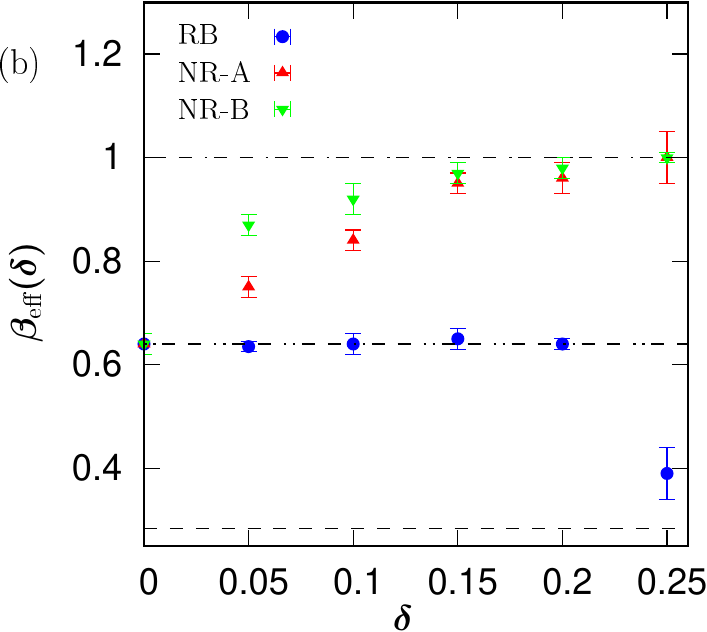} 
	\caption{The effective exponents (a) $\nu_{\text{eff}}$ and (b) $\beta_{\text{eff}}$ for different $\delta$ and bias protocols. Horizontal lines mark the 2D Manna class (``..- -''), the mean-field related values (``.- -''), and exponents for 1D Manna model (``- -''), respectively. Reciprocal bias (RB) preserves the Manna universality class (except $\delta=0.25$ where the system shrink into the 1D conserved Manna model), while both non-reciprocal protocols (NR-A and NR-B) drive a flow to mean-field criticality.}
	\label{manna_exponents}
\end{figure}

For each $\delta$ and protocol, the critical density $\rho_c(\delta)$ is located via the behavior of $U_L$, {\it i.e.}, for the results of $U_L$ with  $L=256$, we plot $\frac{2}{3}-U_L$ {\it v.s.} $\rho-\rho'$. By adjusting $\rho'$ continuously, we obtain a best power-law behavior of $U_L$ at $\rho_c=\rho'$ (Fig. \ref{manna_tc} (a)). 
 Then, we apply the scaling law $\frac{2}{3}-U_L\sim |t|^{-d\nu_{\text{eff}}}$ ($d = 2$) to directly extract the effective correlation length exponent $\nu_{\text{eff}}$. The order parameter scaling $\langle m\rangle\sim |t|^{\beta_{\text{eff}}}$ yields $\beta_{\text{eff}}$ ({\it Supplement Materials S3}).

For the RB protocol, $\rho_c(\delta)$ increases with $\delta$ (FIG. \ref{manna_tc} (b), blue circles). Crucially, the exponents  $\nu_{\text{eff}}$ and $\beta_{\text{eff}}$ remain constant (except to $\delta=0.25$ where the system reduce to $L$-lines of 1D conserved Manna sandpile with random initial conditions, which is believed to belong to the $(1+1)$-dimensional directed percolation universality class \cite{basu2012fixed,kwon2015critical}, with critical exponents $\nu\approx 1.09$ and $\beta\approx 0.28$, repectively.), equal to the established 2D conserved Manna class values ($\nu\approx 0.80$, $\beta\approx 0.64$) \cite{lubeck2003universal} (FIG. \ref{manna_exponents}, blue circles). Therefore, reciprocal anisotropy shifts $\rho_c$ but preserves the universality class. 

In stark contrast, both non-reciprocal protocols (NR-A, NR-B) decrease the critical density (Fig. \ref{manna_tc} (b), red and green symbols) and induce a systematic flow of exponents. For any $\delta>0$, $\nu_{\text{eff}}$ and $\beta_{\text{eff}}$ deviate from Manna class values and flow rapidly with increasing $\delta$ toward the mean-field related values ($\beta_{\text{eff}}\rightarrow \beta_{\text{MF}}\equiv 1$ and $d \nu_{\text{eff}} \rightarrow  d_c \nu_{\text{MF}} \equiv 2$) for conserved Manna-type transitions \cite{karmakar2004directed,henkel2008} (FIG. \ref{manna_exponents}, red and green symbols). Non-reciprocity is thus a relevant perturbation driving the renormalization group flow to the mean-field fixed point.

\begin{figure}[htp]
	\centering
		\includegraphics[width=0.48\linewidth]{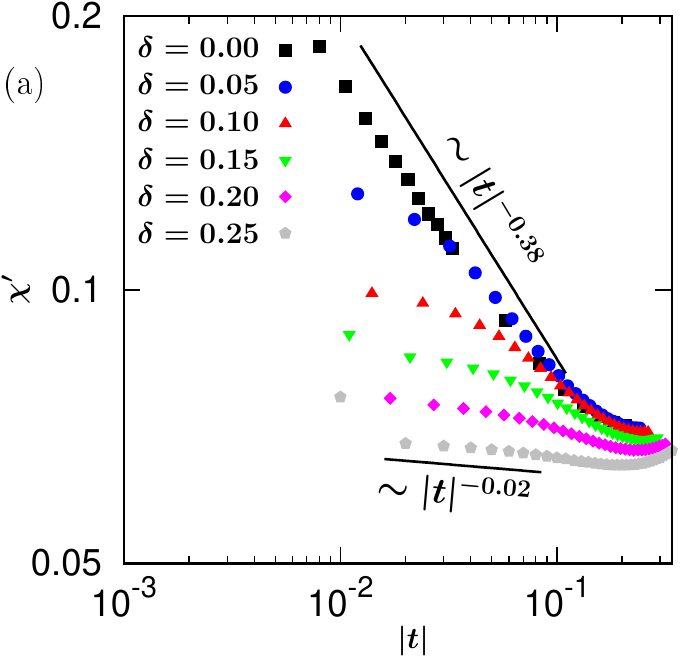}\  \hspace{1mm}
	\includegraphics[width=0.48\linewidth]{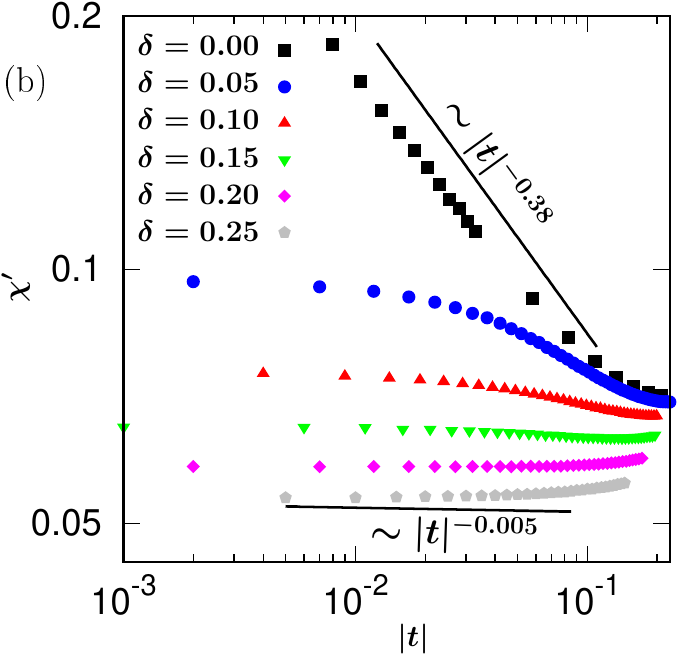} 
	\caption{The log-log plot of $\chi'$ {\it v.s.} $|t|$ for NR-A (a) and NR-B (b) with $L=256$ and different $\delta$. It shows that the exponent of fluctuation $\gamma'$ reduces to $0$ when increasing $\delta$, which indicating that the non-reciprocal interactions suppressing large scale fluctuations.} 
	\label{fig4}
\end{figure}

The dichotomy is now unambiguously established: reciprocal anisotropy (which maintains local detailed balance) preserves the Manna universality class, whereas non-reciprocity (which breaks detailed balance and inversion symmetry) destabilizes it (The results of the 1D and 3D conserved Manna sandpiles with non-reciprocal hopping probabilities confirms this theory, see {\it Supplement Materials S1, S4} for more information.). We posit that the persistent directed current inherent to non-reciprocal dynamics introduces effective long-range spatio-temporal correlations, suppressing large-scale fluctuations. This effectively increases the system's dynamical dimensionality, pushing it toward its upper critical dimension where mean-field theory becomes exact \cite{grinstein1985statistical}. 

To prove this hypothesis, we define the fluctuation of the order parameter as
\begin{equation}
\chi'=N (\langle m^2\rangle - \langle m\rangle^2).
\end{equation}

The scaling behavior of the fluctuation expresses as\cite{lubeck2002scaling}
\begin{equation}
\chi'=|t|^{-\gamma'}.
\end{equation}

Due to the breakdown of the microscopic reciprocity, the fluctuation reduces, and $\gamma'\rightarrow \gamma'_{\text{MF}}\equiv 0$ (Fig. \ref{fig4} (a)-(b)). 


The findings mentioned above has profound implications for non-equilibrium critical phenomena. It suggests that the rich non-mean-field universality classes of isotropic conserved systems are generically unstable to the breaking of reciprocity-a common feature in active matter and driven systems \cite{fruchart2021non,duan2025phase,martin2025transition,belyansky2025phase,
avni2025nonreciprocal,shaebani2020computational}. The rapid crossover implies that modest intrinsic directional driving in experiments may obscure underlying non-mean-field physics, guiding the interpretation of critical exponents in systems from vibrated granules to motile cellular collectives.


{\it Conclusions and discussions--} In summary, this work provides a dual advance in the study of critical phenomena, bridging a methodological gap in finite-size scaling and resolving a key physical question regarding the stability of non-equilibrium universality classes.

Methodologically, we have discovered and rigorously validated a novel scaling regime for the Binder cumulant. By shifting focus from the immediate vicinity of the critical point to the pre-asymptotic region, we established a direct power-law relationship: $U_L\sim N^{-1}|t|^{-d\nu}$ in the disordered phase and 
$\frac{2}{3}-U_L\sim N^{-1} |t|^{-d \nu}$ in the ordered phase. This framework, verified in the 2 and 3D Ising models, 2D site percolation, and at the mean-field fixed point, provides a robust complementary tool for extracting the correlation length exponent $\nu$, especially valuable in systems plagued by strong scaling corrections.

Physically, we deployed this new tool to unravel the impact of microscopic non-reciprocity on the conserved Manna sandpile universality class. Our results reveal a sharp dichotomy. Spatially symmetric, reciprocal biases alter the location of the critical point but leave the universal critical exponents unchanged. In stark contrast, any finite non-reciprocal interaction, whether implemented through structured spatial protocols or completely random asymmetric couplings, acts as a relevant perturbation. It systematically destabilizes the original Manna fixed point, driving the renormalization group flow toward the mean-field universality class. The rapid convergence of the effective exponents $\nu_{\text{eff}}$ and $\beta_{\text{eff}}$ to their mean-field related values demonstrates that non-reciprocity is a potent mechanism for inducing mean-field criticality in conserved, non-equilibrium systems.

The connection between our methodological innovation and physical discovery is profound. The sensitivity of the pre-asymptotic $\frac{2}{3}-U_L$ scaling to the effective exponent $\nu_{\text{eff}}$ was crucial for cleanly diagnosing the continuous flow of the critical properties. This allowed us to distinguish the stable fixed point under reciprocal perturbation from the unambiguous crossover driven by non-reciprocity, a task for which traditional scaling analyses near $T_c$ are often less conclusive.

The physical mechanism behind this flow likely stems from the fundamental symmetry breaking induced by non-reciprocity. The persistent, directional currents break detailed balance and inversion symmetry at the microscopic scale. This can generate effective, long-range spatiotemporal correlations in the avalanche dynamics, suppressing large-scale fluctuations. In renormalization group terms, this is akin to increasing the effective dynamical dimensionality of the system, thereby pushing it toward its upper critical dimension where mean-field theory becomes exact \cite{grinstein1985statistical}. The finding that even weak non-reciprocity induces this flow suggests that the well-studied non-mean-field universality classes of isotropic, conserved systems may be surprisingly fragile in the broader context of active and driven matter, where non-reciprocal interactions are the rule rather than the exception.

Our work opens several promising avenues for future research. First, the new Binder cumulant scaling law should be tested in other challenging contexts, such as disordered systems or models with multiple competing order parameters, where traditional FSS struggles. Second, the generality of the ``non-reciprocity-induced mean-field'' mechanism should be explored in other non-equilibrium universality classes, such as directed percolation or the Kardar-Parisi-Zhang equation under conservative driving. 

Ultimately, this study enhances our toolbox for analyzing criticality and provides a unifying principle for understanding phase transitions in a wide array of intrinsically non-equilibrium systems, from synthetic active materials to biological networks.

\section*{Acknowlegments}

We thank the helpful discussions with S. Fang. WZ acknowledges support by the National Natural Science Foundation of China Youth Fund (under Grant No. 12105133), and the start-up fund of Minjiang University. YD is supported by the National Natural Science Foundation of China (Grant No. 12275263) and the Natural Science Foundation of Fujian Province of China (Grant No. 2023J02032).

\bibliography{manna}

\onecolumngrid
\newpage
\section*{Supplement Materials for ``Crossover Scaling of Binder Cumulant and its application in Non-reciprocal Sandpiles''}
\begin{center}
Wei Zhong$^1$, Youjin Deng$^{2,3}$ $^\star$ \\
yjdeng@ustc.edu.com\\
$^1$ Minjiang Collaborative Center for Theoretical Physics, College of Physics and Electronic Information Engineering,
Minjiang University, Fuzhou 350108, P. R. China.\\
$^2$ Hefei National Laboratory for Physical Sciences at the Microscale and Department of Modern Physics, University of Science and Technology of China, Hefei 230026, China\\
$^3$ Hefei National Laboratory, University of Science and Technology of China, Hefei 230088, China
\end{center}





This Supplemental Material contains 1) the simulation results of the Binder cumulant for the 3D Ising model and 2D site percolation, 2) mean-filed theory of the Binder cumulant for the Ising model, 3) data for extracting the critical exponents for the 2D conserved Manna model with non-reciprocal interactions, and 4) critical exponents for conserved Manna model with non-reciprocal interactions and different spatial dimensions.

\section*{S1. Scaling behavior of the Binder cumulant for the 3D Ising model and 2D site percolation}
 
\begin{figure}[htp]
	\centering
	\includegraphics[width=0.6\linewidth]{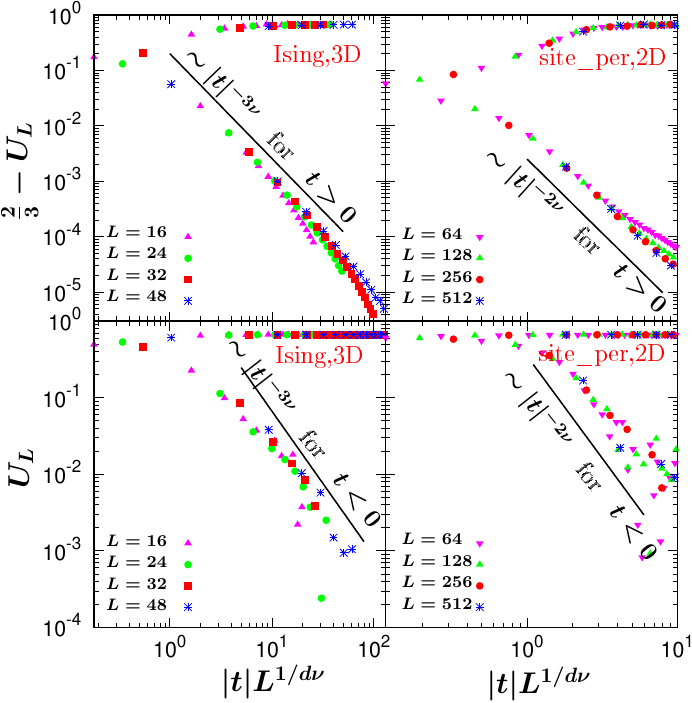} 
	\caption{Log-log plot of $\frac{2}{3}-U_L$ and $U_L$ {\it v.s.} $|t| L^{1/d\nu}$ for three-dimensional Ising model (Left) and 2D site percolation system (Right). }
	\label{fig_3dising_siteper}
\end{figure}

 To further confirm the scaling behavior of the Binder cumulant near the critical point, we calculate the Binder cumulant for the 3D Ising model and 2D site percolation near the critical points. The simulation result depicted in Fig. \ref{fig_3dising_siteper} confirms the scaling behavior reported in Eq. (2)-(3) of the main tex.
 
\section*{S2. Mean-filed theory of the Binder cumulant for the Ising model} 

We compute the Binder cumulant \(U = 1 - \langle m^4 \rangle / (3 \langle m^2 \rangle^2)\) for the mean-field Ising model by direct saddle-point expansion of the moment integrals. The probability distribution is
\begin{equation}
P(m) \propto e^{-N \beta f(m)}, \quad f(m) = \frac{a}{2}m^2 + \frac{b}{4}m^4,
\end{equation}
where \(a = a_0(T - T_c)\), \(b > 0\) is constant, \(\beta = 1/(k_B T)\), and \(N\) is the total number of spins.

\subsection{General method}

The moments are computed via
\begin{equation}
\langle m^k \rangle = \frac{\int_{-\infty}^{\infty} m^k e^{-N\beta f(m)} \, dm}{\int_{-\infty}^{\infty} e^{-N\beta f(m)} \, dm}.
\end{equation}
For large \(N\), we apply saddle-point expansion around the minima of \(f(m)\). Let \(m_*\) be a minimum (\(f'(m_*) = 0\)). Expand:
\begin{equation}
f(m) \approx f(m_*) + \frac{1}{2}f''(m_*)(m - m_*)^2 + \frac{1}{6}f'''(m_*)(m - m_*)^3 + \frac{1}{24}f^{(4)}(m_*)(m - m_*)^4 + \cdots
\end{equation}
For our quartic \(f\), \(f'''(m) = 0\) and \(f^{(4)}(m) = 6b\). The Gaussian integrals are corrected by the quartic term via perturbation expansion.

\subsection{High-temperature phase (\(T > T_c\))}

Minimum at \(m_* = 0\), with \(f(0) = 0\), \(f''(0) = a > 0\), \(f^{(4)}(0) = 6b\). The partition function:
\begin{equation}
Z = \int_{-\infty}^{\infty} e^{-N\beta \left[ \frac{a}{2}m^2 + \frac{b}{4}m^4 \right]} dm.
\end{equation}

Expand the quartic term: \(e^{-N\beta b m^4/4} = 1 - \frac{N\beta b}{4}m^4 + O(N^{-2})\). Then:
\begin{equation}
Z = \int_{-\infty}^{\infty} e^{-N\beta a m^2/2} \left[ 1 - \frac{N\beta b}{4}m^4 \right] dm + O(N^{-1}).
\end{equation}

Using Gaussian integrals:
\begin{equation}
Z = \sqrt{\frac{2\pi}{N\beta a}} \left[ 1 - \frac{3b}{4N\beta a^2} \right] + O(N^{-2}).
\end{equation}

Similarly,
\begin{align}
\int m^2 e^{-N\beta f(m)} dm = \int m^2 e^{-N\beta a m^2/2} \left[ 1 - \frac{N\beta b}{4}m^4 \right] dm = \sqrt{\frac{2\pi}{N\beta a}} \left[ \frac{1}{N\beta a} - \frac{15b}{4(N\beta a)^2} \right],\\
\int m^4 e^{-N\beta f(m)} dm = \int m^4 e^{-N\beta a m^2/2} \left[ 1 - \frac{N\beta b}{4}m^4 \right] dm = \sqrt{\frac{2\pi}{N\beta a}} \left[ \frac{3}{(N\beta a)^2} - \frac{105b}{4(N\beta a)^3} \right].
\end{align}

Dividing by \(Z\) to obtain moments:
\begin{equation}
\langle m^2 \rangle = \frac{1}{N\beta a} - \frac{3b}{(N\beta a)^2} + O(N^{-3}), \quad \langle m^4 \rangle = \frac{3}{(N\beta a)^2} - \frac{24b}{(N\beta a)^3} + O(N^{-4}).
\end{equation}
 
Thus:
\begin{equation}
U = 1 - \frac{\langle m^4 \rangle}{3\langle m^2 \rangle^2} = 1 - \frac{ \frac{3}{(N\beta a)^2} - \frac{24b}{(N\beta a)^3} }{ 3\left[ \frac{1}{(N\beta a)^2} - \frac{6b}{(N\beta a)^3} \right] } + O(N^{-2}) = \frac{2b}{N\beta a^2} + O(N^{-2}).
\end{equation}

\subsection{Critical point (\(T = T_c\))}

Here \(a = 0\), \(f(m) = \frac{b}{4}m^4\). Exact computation:
\begin{equation}
Z = \int_{-\infty}^{\infty} e^{-N\beta b m^4/4} dm = 2 \int_0^{\infty} e^{-\alpha m^4} dm, \quad \alpha = \frac{N\beta b}{4}.
\end{equation}
 
Let \(u = \alpha m^4\), then \(dm = \frac{1}{4}\alpha^{-1/4} u^{-3/4} du\), so
\begin{equation}
Z = \frac{1}{2}\alpha^{-1/4} \Gamma(1/4).
\end{equation}

Similarly,
\begin{align}
\langle m^2 \rangle = \frac{1}{Z} \int m^2 e^{-\alpha m^4} dm = \frac{ \frac{1}{2}\alpha^{-3/4} \Gamma(3/4) }{ \frac{1}{2}\alpha^{-1/4} \Gamma(1/4) } = \alpha^{-1/2} \frac{\Gamma(3/4)}{\Gamma(1/4)},\\
\langle m^4 \rangle = \frac{1}{Z} \int m^4 e^{-\alpha m^4} dm = \frac{ \frac{1}{2}\alpha^{-1} \Gamma(5/4) }{ \frac{1}{2}\alpha^{-1/4} \Gamma(1/4) } = \alpha^{-1} \frac{\Gamma(5/4)}{\Gamma(1/4)} = \frac{1}{4\alpha},
\end{align}
 
since \(\Gamma(5/4) = \frac{1}{4}\Gamma(1/4)\). Thus:
\begin{equation}
U = 1 - \frac{1/(4\alpha)}{3 \left[ \alpha^{-1} \left( \frac{\Gamma(3/4)}{\Gamma(1/4)} \right)^2 \right]} = 1 - \frac{1}{12} \left[ \frac{\Gamma(1/4)}{\Gamma(3/4)} \right]^2 \approx 0.27044.
\end{equation}

This result is consistent with the mean-field value of Binder cumulant obtained in Ref. [1]

\subsection{Low-temperature phase (\(T < T_c\))}

Here \(a < 0\). Two minima at \(m = \pm m_0\), \(m_0 = \sqrt{-a/b}\). Near \(m_0\), let \(m = m_0 + \delta\), then
\begin{equation}
f(m_0 + \delta) = f(m_0) + \frac{1}{2}\kappa \delta^2 + \frac{b}{2}m_0 \delta^3 + \frac{b}{4}\delta^4, \quad \kappa = f''(m_0) = -2a > 0.
\end{equation}

The partition function is the sum of contributions from both minima. For one minimum:
\begin{equation}
Z_+ = \int e^{-N\beta [f(m_0) + \frac{1}{2}\kappa \delta^2 + \frac{b}{2}m_0 \delta^3 + \frac{b}{4}\delta^4]} d\delta.
\end{equation}

Expand the exponent: \(e^{-N\beta \frac{b}{2}m_0 \delta^3} = 1 - N\beta \frac{b}{2}m_0 \delta^3 + \cdots\), and \(e^{-N\beta \frac{b}{4}\delta^4} = 1 - N\beta \frac{b}{4}\delta^4 + \cdots\). The cubic term averages to zero in Gaussian integration. Keeping terms up to \(O(1/N)\):
\begin{equation}
Z_+ = e^{-N\beta f(m_0)} \int e^{-N\beta \kappa \delta^2/2} \left[ 1 - N\beta \frac{b}{4}\delta^4 \right] d\delta = e^{-N\beta f(m_0)} \sqrt{\frac{2\pi}{N\beta \kappa}} \left[ 1 - \frac{3b}{4N\beta \kappa^2} \right].
\end{equation}

The total \(Z = 2Z_+\) (both minima). For the moments, note \(\langle m^k \rangle = \frac{1}{Z} \left[ \int_{m_0} m^k e^{-N\beta f} + \int_{-m_0} m^k e^{-N\beta f} \right]\). For even \(k\), both minima contribute equally. Compute for one minimum (around \(m_0\)):
\begin{equation}
\int m^2 e^{-N\beta f} dm = e^{-N\beta f(m_0)} \int (m_0^2 + 2m_0\delta + \delta^2) e^{-N\beta [\frac{1}{2}\kappa \delta^2]} \left[ 1 - N\beta \frac{b}{4}\delta^4 \right] d\delta.
\end{equation}

The linear term in \(\delta\) vanishes. To \(O(1/N)\):
\begin{equation}
\int m^2 e^{-N\beta f} dm = e^{-N\beta f(m_0)} \left[ m_0^2 \sqrt{\frac{2\pi}{N\beta \kappa}} \left(1 - \frac{3b}{4N\beta \kappa^2}\right) + \sqrt{\frac{2\pi}{N\beta \kappa}} \left( \frac{1}{N\beta \kappa} - \frac{3b}{4(N\beta \kappa)^2} \right) \right].
\end{equation}

Thus the contribution from one minimum is:
\begin{equation}
e^{-N\beta f(m_0)} \sqrt{\frac{2\pi}{N\beta \kappa}} \left[ m_0^2 + \frac{1}{N\beta \kappa} - \frac{3b m_0^2}{4N\beta \kappa^2} - \frac{3b}{4(N\beta \kappa)^2} \right].
\end{equation}

Doubling for two minima and dividing by \(Z = 2 e^{-N\beta f(m_0)} \sqrt{\frac{2\pi}{N\beta \kappa}} \left[ 1 - \frac{3b}{4N\beta \kappa^2} \right]\), we obtain:
\begin{equation}
\langle m^2 \rangle = m_0^2 + \frac{1}{N\beta \kappa} + O(N^{-2}).
\end{equation}
 
Similarly,
\begin{equation}
\int m^4 e^{-N\beta f} dm \text{ (one minimum)} = e^{-N\beta f(m_0)} \int (m_0^4 + 4m_0^3\delta + 6m_0^2\delta^2 + 4m_0\delta^3 + \delta^4) e^{-N\beta \kappa \delta^2/2} \left[ 1 - N\beta \frac{b}{4}\delta^4 \right] d\delta.
\end{equation}

Only even powers in \(\delta\) survive. To \(O(1/N)\):
\begin{equation}
= e^{-N\beta f(m_0)} \sqrt{\frac{2\pi}{N\beta \kappa}} \left[ m_0^4 + 6m_0^2 \left( \frac{1}{N\beta \kappa} \right) + 3 \left( \frac{1}{N\beta \kappa} \right)^2 - \frac{3b m_0^4}{4N\beta \kappa^2} - \frac{18b m_0^2}{4N\beta \kappa^3} - \frac{9b}{4N\beta \kappa^4} \right].
\end{equation}

Averaging over both minima and dividing by \(Z\):
\begin{equation}
\langle m^4 \rangle = m_0^4 + 6m_0^2 \frac{1}{N\beta \kappa} + 3 \left( \frac{1}{N\beta \kappa} \right)^2 + O(N^{-2}).
\end{equation}

Now \(\kappa = -2a\), \(m_0^2 = -a/b\). Thus:
\begin{equation}
\langle m^2 \rangle = m_0^2 + \sigma^2, \quad \langle m^4 \rangle = m_0^4 + 6m_0^2 \sigma^2 + 3\sigma^4,
\end{equation}

with \(\sigma^2 = 1/(N\beta \kappa) = 1/(N\beta (-2a))\). Then:
\begin{equation}
U = 1 - \frac{m_0^4 + 6m_0^2\sigma^2 + 3\sigma^4}{3(m_0^2 + \sigma^2)^2} = \frac{2}{3} - \frac{4\sigma^2}{3m_0^2} + O(N^{-2}) = \frac{2}{3} - \frac{2b}{3N\beta a^2} + O(N^{-2}),
\end{equation}

since \(\sigma^2/m_0^2 = \frac{1/(N\beta (-2a))}{(-a/b)} = \frac{b}{2N\beta (-a)^2} = \frac{b}{2N\beta a^2}\) (note \(a<0\), so \(a^2 = (-a)^2\)).

\subsection{Summary of results}

\begin{center}
\begin{tabular}{@{}ccc@{}}
\toprule
Regime & \(U\) up to \(O(1/N)\) & Thermodynamic limit \(U\) \\
\(T > T_c\) & \(\displaystyle \frac{2b}{N\beta a^2}\) & \(0\) \\
\(T = T_c\) & \(\displaystyle 1 - \frac{1}{12} \left[ \frac{\Gamma(1/4)}{\Gamma(3/4)} \right]^2 \approx 0.27044\) & \(0.27044\) \\
\(T < T_c\) & \(\displaystyle \frac{2}{3} - \frac{2b}{3N\beta a^2}\) & \(\displaystyle \frac{2}{3}\) \\
\end{tabular}
\end{center}
where \(a = a_0(T - T_c)\), \(\beta = 1/(k_B T)\), and \(b\) is proportional to $T$ for the Ising model [2].

\section*{S3. Data for extracting the critical exponents for the 2D conserved Manna model with non-reciprocal interactions}

Near the critical points of the conserved Manna sandpile, we expect the following scaling behavior of the Binder cumulant $U_L$, the order parameter $m$, and the fluctuation $\chi'$ as
\begin{align*}
\frac{2}{3}-U_L &\sim (|\rho-\rho_c| L^{1/\nu})^{-d\nu}\\
\langle m\rangle\ L^{\beta/\nu} &\sim (|\rho-\rho_c| L^{1/\nu})^{\beta}\\
\chi' &\sim (|\rho-\rho_c| )^{-\gamma'}
\end{align*}

These relations provides us with the values of the exponents of $\nu$, $\beta$, and $\gamma'$ independently.

\begin{figure}[htp]
	\centering
	\includegraphics[width=0.31 \linewidth]{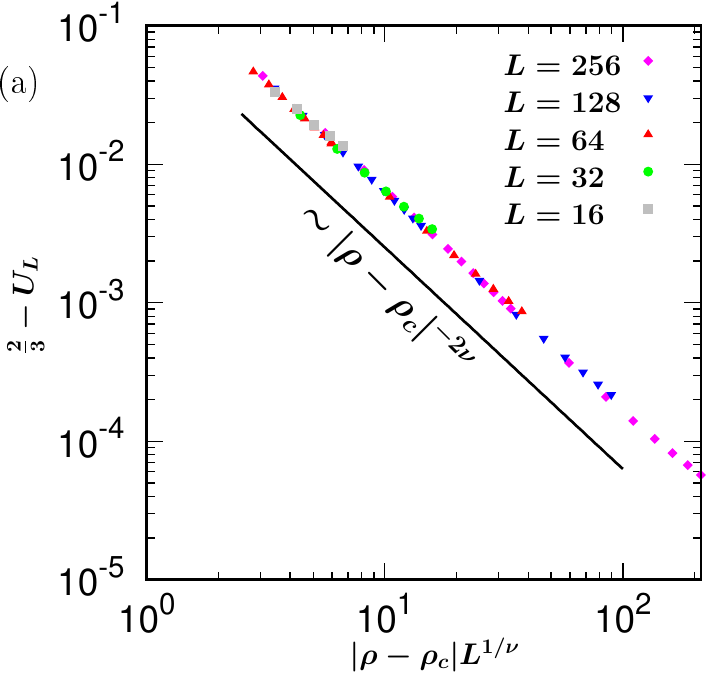} 
	\hspace{1mm}
	\includegraphics[width=0.31 \linewidth]{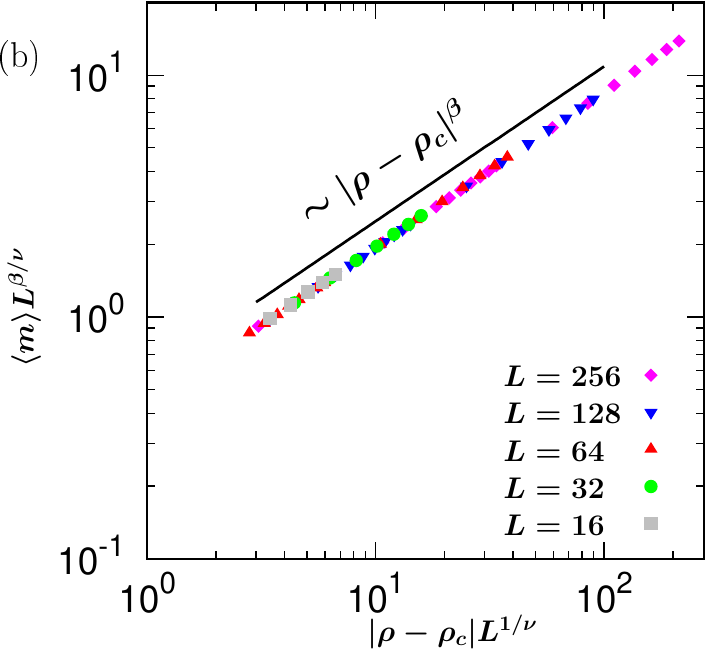} \hspace{1mm}
	\includegraphics[width=0.31\linewidth]{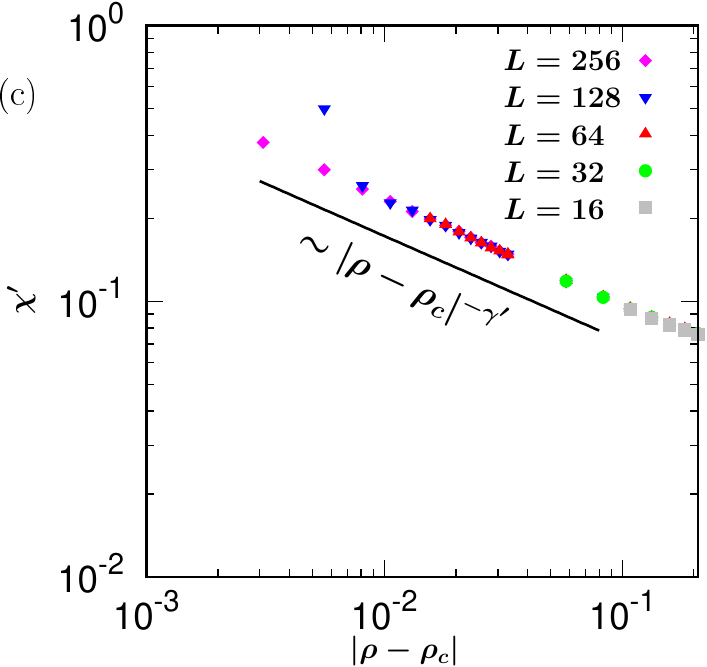} 
	\caption{(a)-(c) The scaling behavior of $\frac{2}{3}-U_L$, $\langle m\rangle$, and $\chi'$ for the pure 2D conserved Manna sandpile, respectively. }
	\label{2d_cms}
\end{figure}

Fig. \ref{2d_cms} shows that for the pure 2D conserved Manna sandpile, we extract the exponents from the scaling behavior and we find $\nu=0.80(2)$, $\beta=0.64(2)$ and $\gamma'=0.38(2)$ which are consistent with the results reported in Ref. [3]

 In this paper, we consider three different bias to the system. They are 
 
 {\bf 1 Reciprocal Bias (RB, Case I)}: $q_{\text{right}}=q_{\text{left}}=0.25+\delta$, $q_{\text{up}}=q_{\text{down}}=0.25-\delta$. This preserves spatial inversion symmetry. 

{\bf 2 Non-reciprocal Bias A (NR-A, Case II)}: $q_{\text{right}}=q_{\text{left}}=q_{\text{up}}=0.25+\delta/3$, $q_{\text{down}}=0.25-\delta$.

{\bf 3 Non-reciprocal Bias B (NR-B, Case III)}: $q_{\text{right}}=q_{\text{up}}=0.25+\delta$, $q_{\text{left}}=q_{\text{down}}=0.25-\delta$. 

 \begin{figure}[htp]
	\centering
	\includegraphics[width=0.31\linewidth]{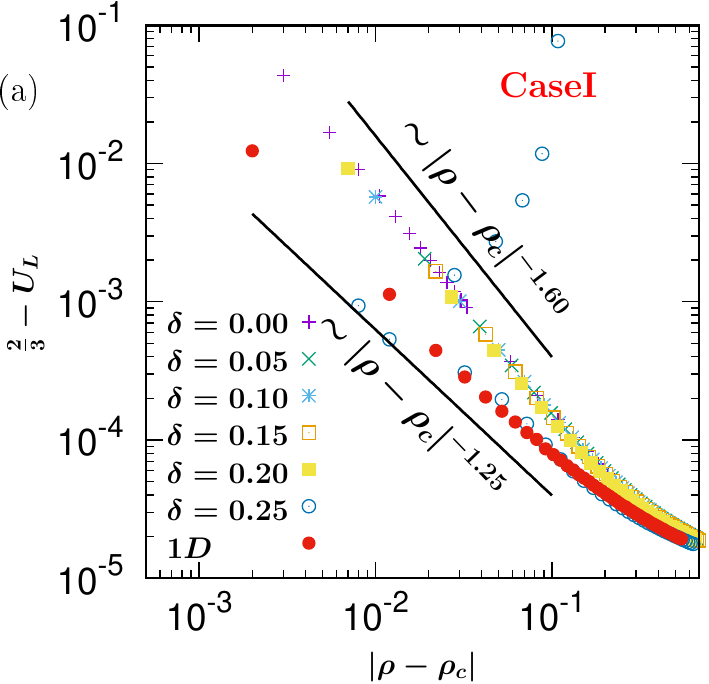} 
	\hspace{1mm}
	\includegraphics[width=0.31\linewidth]{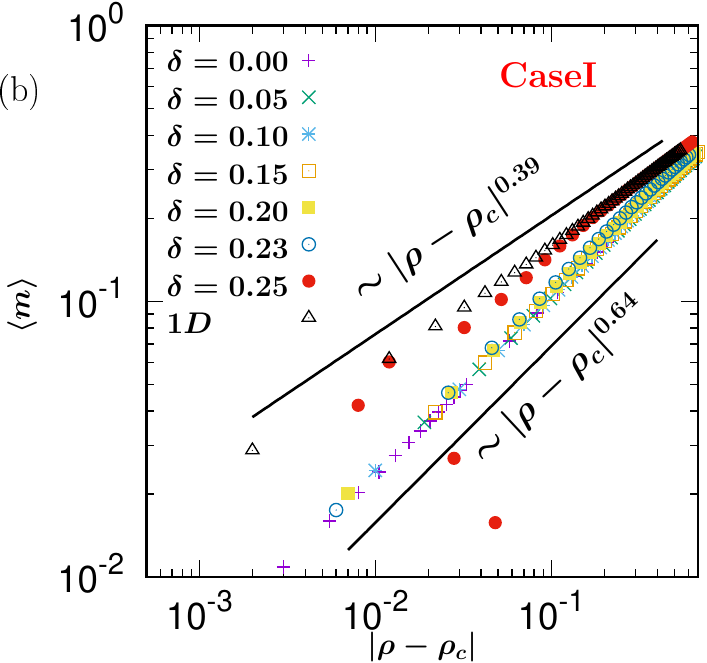} \hspace{1mm}
	\includegraphics[width=0.31\linewidth]{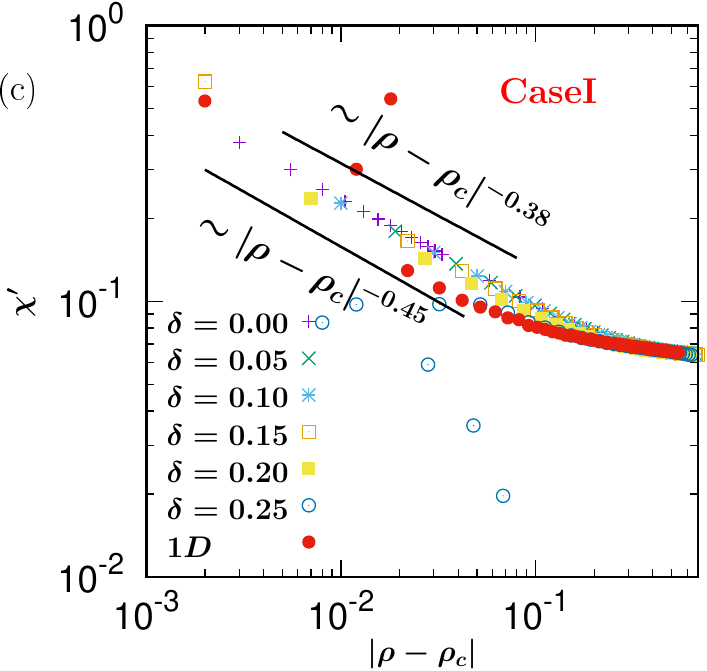} 
	\caption{(a)-(c) Critical behavior of $\frac{2}{3}-U_L$, $\langle m\rangle$, and $\chi'$ for {\color{blue}{$RB (Case I) $}} with different $\delta$ and $L=256$, respectively. It shows that except $\delta=0.25$, the reciprocal bias interaction preserve the universality class with the standard conserved Manna sandpile. For $\delta=0.25$, it reduces to the behavior of the 1D conserved Manna sandpile.}
	\label{case1}
\end{figure}

For the reciprocal bias, the result in Fig. \ref{case1} shows that for different $\delta<0.25$, the behavior of the Binder cumulant, order parameter, and fluctuation are consistence with the result for the pure conserved Manna sandpile, which means reciprocal bias keeps the universality class of the system. 

\begin{figure}[htp]
	\centering
	\includegraphics[width=0.31 \linewidth]{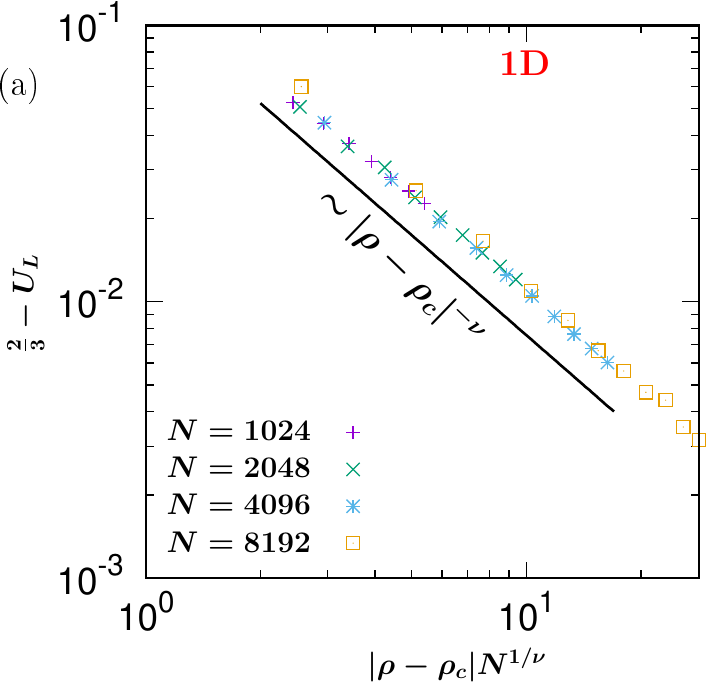} 
	\hspace{1mm}
	\includegraphics[width=0.31 \linewidth]{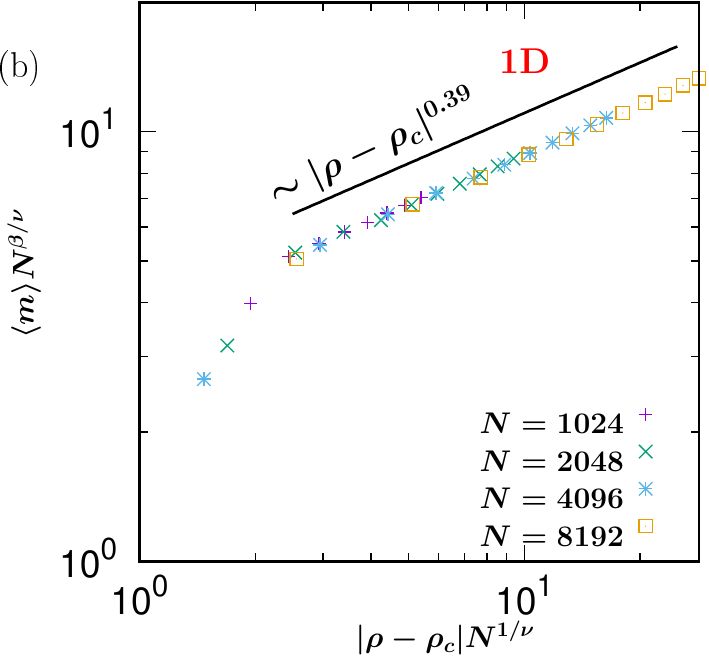}
	 \hspace{1mm} 
	\includegraphics[width=0.31 \linewidth]{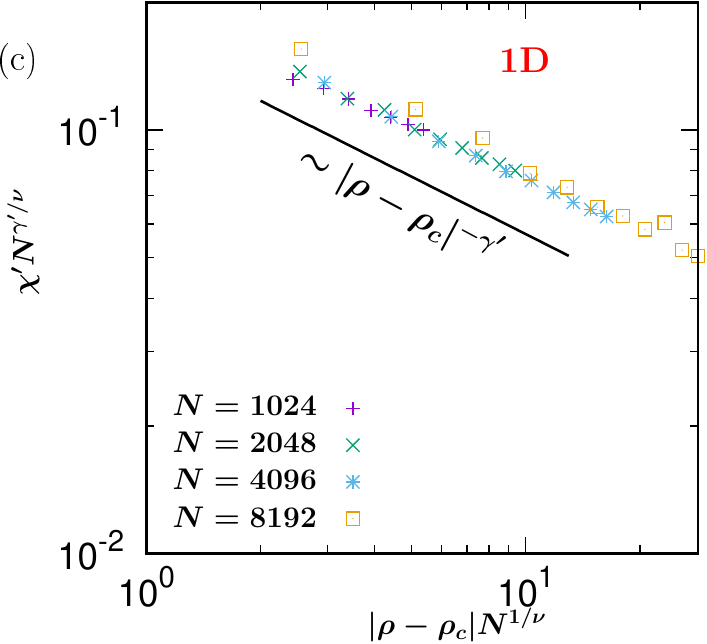}
	\caption{The critical behavior of the 1D conserved Manna sandpile. The obtained critical exponents are consistence with the exits results [3].}
	\label{1d_manna}
\end{figure}

However, we should note that for $\delta=0.25$, the system reduces to the 1D conserved Manna model with randomly distributed initial condition. Therefore, when $L\rightarrow \infty$, it shares the same critical exponents as the 1D Manna sandpile, as shown in Fig. \ref{1d_manna}. Here, we achieve $\nu=1.15(3)$, $\beta=0.39(1)$, and $\gamma=0.45(2)$, which is in consistence with the results in Ref.[3].

\begin{figure}[htp]
	\centering
	\includegraphics[width=0.31\linewidth]{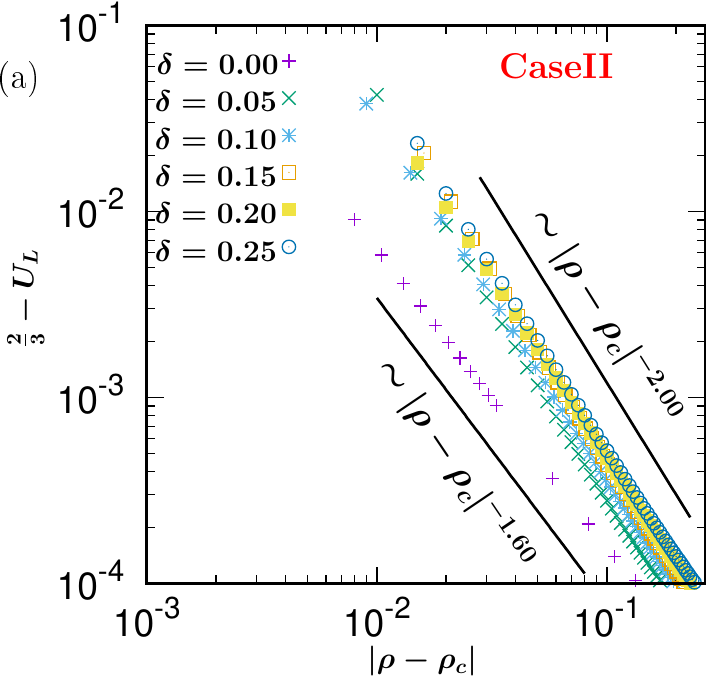} 
	\hspace{2mm}
	\includegraphics[width=0.31\linewidth]{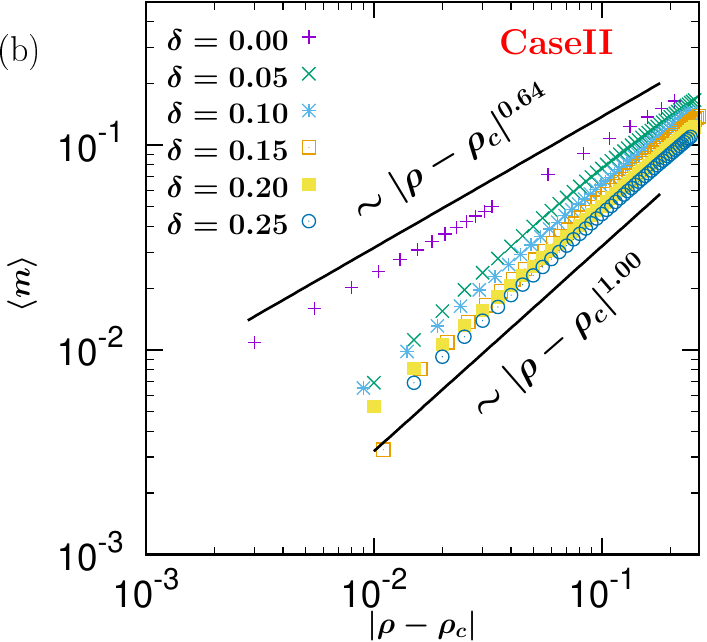} \hspace{1mm}
	\includegraphics[width=0.31\linewidth]{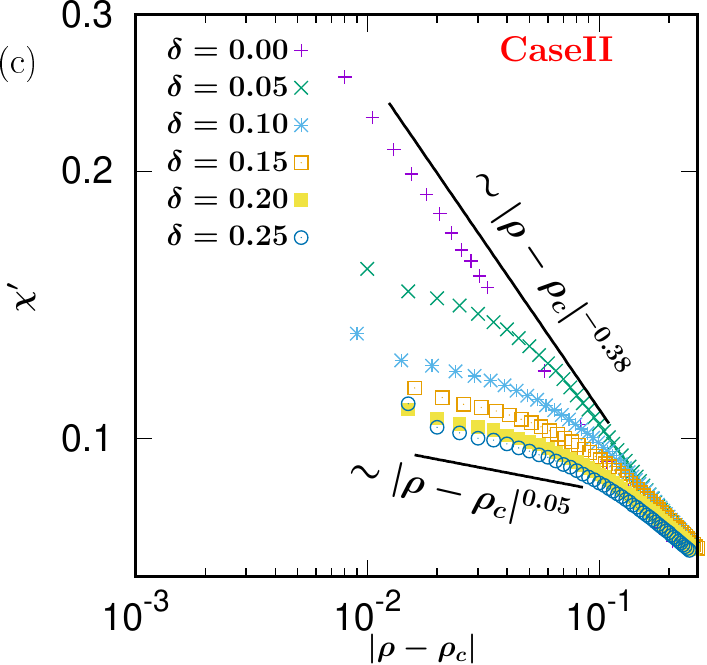}
	\caption{The critical behavior of the non-reciprocal bias interaction ({\color{blue}{NR-A, Case II}}). The results indicate that the non-reciprocal bias drives the system to the mean-field behavior with the critical exponents $\nu=1$, $\beta=1$, and $\gamma'=0$, respectively. }
	\label{case2}
\end{figure}

\begin{figure}[htp]
	\centering
	\includegraphics[width=0.31\linewidth]{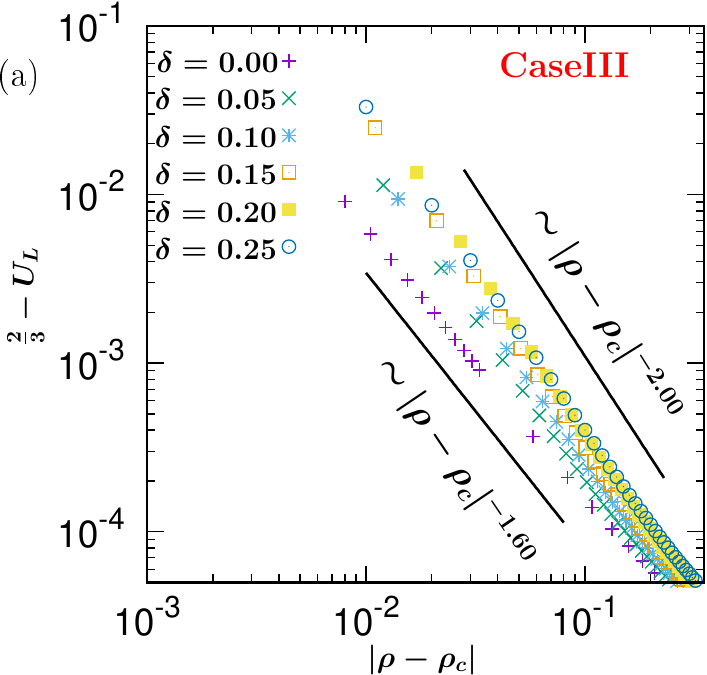} 
	\hspace{2mm}
	\includegraphics[width=0.31\linewidth]{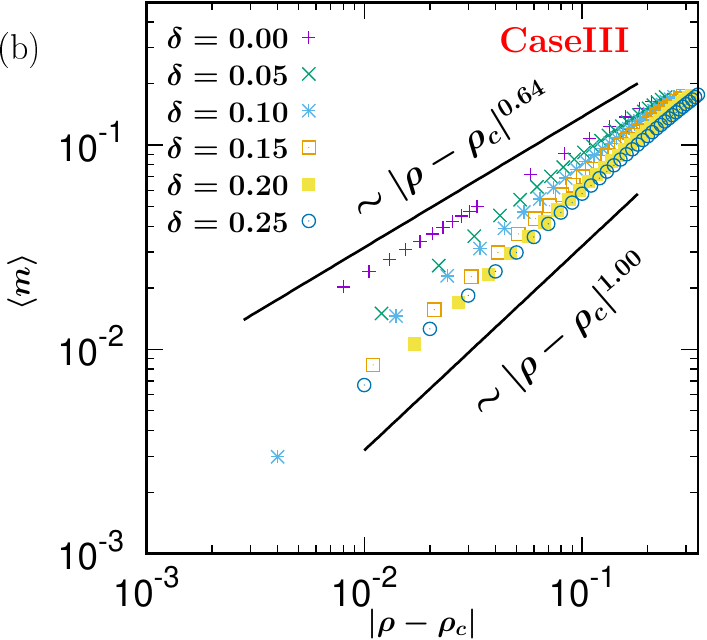} \hspace{1mm}
	\includegraphics[width=0.31\linewidth]{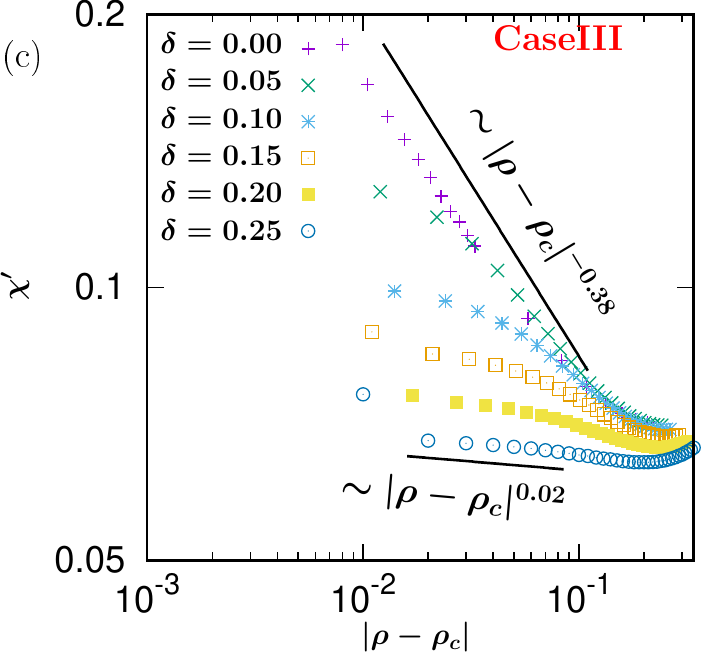}
	\caption{The critical behavior of the non-reciprocal bias interaction ({\color{blue}{NR-B, Case III}}). The results indicate that the non-reciprocal bias ({\color{blue}{Case III}}) also drives the system to the mean-field behavior.}
	\label{case3}
\end{figure}

For non-reciprocal biases, figs. \ref{case2} and \ref{case3} illustrate that even slight non-reciprocal bias drives the system toward the mean-field values of the critical exponents.

\section*{S4. Critical exponents for conserved Manna model with non-reciprocal interactions} 

To further confirm the effect of the non-reciprocal interactions to the critical behavior of conserved Manna sandpile, we explore the 1D and 3D conserved Manna sandpile with non-reciprocal hoping probabilities.

For the 1D case, the hopping rates are expresses as
\begin{align}
p_{\text{right}} & =0.5+\delta \\
p_{\text{left}} & =0.5-\delta
\end{align}
with $\delta$ varies from $0$ to a half.

For the 3D system, we only consider the following situation:
\begin{align*}
p_{\text{right}}=p_{\text{up}}=p_{\text{front}} & =1/6+\delta \\
p_{\text{left}}=p_{\text{down}}=p_{\text{back}} & =1/6-\delta
\end{align*}
 
\begin{figure}[htp]
	\centering
	\includegraphics[width=0.61\linewidth]{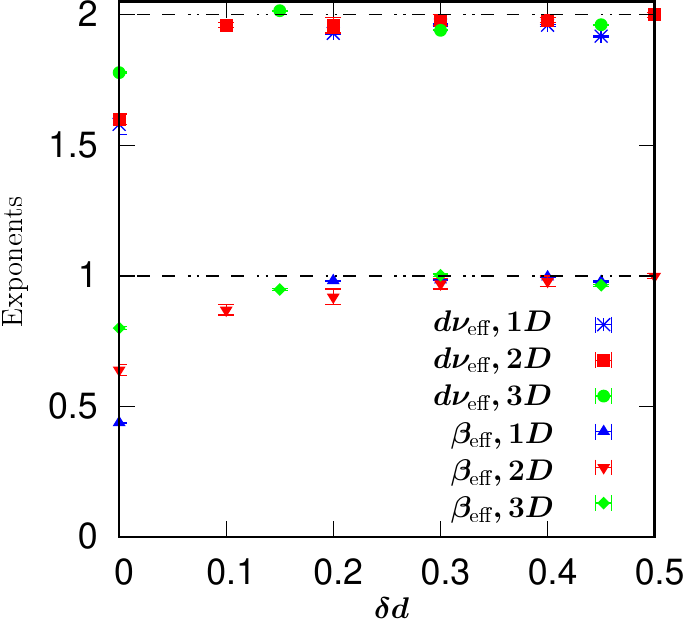} 
	\caption{Critical exponents $d \nu_{\text{eff}}$, $\beta_{\text{eff}}$ for one- to three- dimensional conserved Manna sandpiles with non-reciprocal hoping probabilities. The results confirms that the non-reciprocal interaction drives the conserved Manna sandpile to the mean-field values of critical exponents.}
	\label{fig_s3}
\end{figure}

Combining the recults of the critical exponents for 2D conserved Manna system ($NR-B$), our result (Fig. \ref{fig_s3}) confirms that non-reciprocal interaction drive the conserved Manna sandpile to mean-field critical behavior ($d_c\nu_{\text{MF}}=2$,and $\beta_{\text{MF}}=1$).
 
\section*{S5 References}

[1] Luijten, E. (1997). Interaction range, universality and the upper critical dimension.

[2] A. Wipf, Statistical approach to quantum field theory (Springer, 2021).

[3] S. L\"{u}beck and P. Heger, Universal finite-size scaling behavior and universal dynamical scaling behavior of absorbing phase transitions with a conserved field, Physical Review E 68 , 056102 (2003).
\end{document}